\documentclass[12pt,preprint]{aastex}

\shorttitle{Discovery of a Probable Triple Quasar}
\shortauthors{Djorgovski et al.}

\begin{document}

\title{Discovery of a Probable Physical Triple Quasar\altaffilmark{1}}

\author{S. G. Djorgovski\altaffilmark{2,3},
F. Courbin\altaffilmark{3},
G. Meylan\altaffilmark{3},
D. Sluse\altaffilmark{3},
D. J. Thompson\altaffilmark{4},
A. Mahabal\altaffilmark{2},
E. Glikman\altaffilmark{2}
}

\altaffiltext{1}
{Based in part on the data obtained at the W.M. Keck Observatory, which is operated as a scientific partnership among the California Institute of Technology, the University of California and NASA, made possible by the generous financial support of the W.M. Keck Foundation; and data obtained with VLT/ANTU telescope at ESO-Paranal Observatory; and data obtained with the Hubble Space Telescope, operated by NASA.  This research also made use of data obtained from the US National Virtual Observatory, which is sponsored by the National Science Foundation.}

\altaffiltext{2}{Division of Physics, Mathematics, and Astronomy, California Institute of Technology, Pasadena, CA 91125, USA.}
\altaffiltext{3}{Laboratoire d'Astrophysique, Ecole Polytechnique F\'ed\'erale de Lausanne (EPFL), CH-1290 Sauverny, Switzerland.}
\altaffiltext{4}{Large Binocular Telecope Observatory, Univ. of Arizona, Tucson, AZ 85721, USA.}


\begin{abstract}
We report the discovery of the first known probable case of a physical triple quasar (not a gravitational lens).  A previously known double system, QQ $1429-008$ at $z = 2.076$, is shown to contain a third, fainter QSO component at the same redshift within the measurement errors.  Deep optical and IR imaging at the Keck and VLT telescopes has failed to reveal a plausible lensing galaxy group or a cluster, and moreover, we are unable to construct any viable lensing model which could lead to the observed distribution of source positions and relative intensities of the three QSO image components.  Furthermore, there are hints of differences in broad-band spectral energy distributions of different components, which are more naturally understood if they are physically distinct AGN.  Therefore, we conclude that this system is most likely a physical triple quasar, the first such close QSO grouping known at any redshift.  The projected component separations in the restframe are $\sim 30 - 50$ kpc for the standard concordance cosmology, typical of interacting galaxy systems.  The existence of this highly unusual system supports the standard picture in which galaxy interactions lead to the onset of QSO activity.
\end{abstract}



\section{Introduction}

It is now generally accepted that the onset and fueling of AGN activity is closely related to dissipative tidal interactions and mergers of galaxies, and that there is a close synergy between the formation and growth of supermassive black holes and their host galaxies themselves.  For recent reviews and further references, see, e.g., the proceedings edited by \cite{ho04}, or the study by \cite{hop06}.

If galaxy interactions are responsible for triggering of QSO activity, it is reasonable to expect that in some cases this would occur in both components of an interacting galaxy pair, resulting in a binary QSO.  Starting from the early discoveries of such systems \citep{djo87,mey90}, there are now many tens, if not hundreds of physical binary QSOs known \citep{hen06,mye06}.  Their frequency may be up to two orders of magnitude higher than what may be expected from simple extrapolations of galaxy clustering at comoving scales $< 100$ kpc, which can be understood if interactions enhance the probability of QSO activity, atop of the normal clustering of their host galaxies \citep{djo91,kfm99,hen06}; but the overall picture is still very complex \citep{mye06}.  In many cases, there are ambiguities between the binary QSO and gravitational lensing interpretations of close QSO pairs; see, e.g., \cite{kfm99} or \cite{mwf99} for discussions and references.

One such case is the QSO pair B1429--008 = J1432--0106 \citep{hew89}, whose brighter component was originally found in the LBQS survey \citep{hew91}.  The two brightest QSO components are at $z = 2.076$ and are separated by 5.14 arcsec, corresponding to a projected separation of 43 kpc (proper) or 132 kpc (comoving); we use the standard Friedmann-Lemaitre cosmology with $h = 0.7$, $\Omega_m = 0.3$, and $\Omega_\Lambda = 0.7$ throughout.

In this $Letter$ we report on the discovery of a third QSO component in this system, and discuss additional evidence against the gravitational lensing hypothesis.  We then conclude that this is most likely the first known case of a physical triple QSO, which we denote as QQQ~$1432-0106$ below.

\section{The Data and the Observations}

We obtained images of the field in the Cousins $I$ band, using the LRIS instrument \citep{oke95} at the W.M.  Keck Observatory's 10-m telescope on 13 April 1994 UT, in good conditions.  Eight exposures of 300 s each were obtained, and processed using standard techniques.  A section of the coadded image is shown in Fig. 1.  Several faint sources surround the previously known QSO pair.  We also obtained $K$ band images using the NIRC instrument \citep{ms94} on 06 April 1994 UT, and again on 15 June 1995 UT; the total usable exposure time was 3360 s.  For the purposes of the present analysis, we coadded these data with the ESO VLT $K_s$ band images described by \cite{fau03}.  The central portion of the stacked image is shown in Fig. 2.  We use the same notation for the fainter sources as in \cite{fau03}, except for their source \#5, which we label as C.  The limiting magnitudes (1 $\sigma$) for the stacked $K$ band image are 24.80, 24.05, and 23.30 mag for the apertures of 1, 2, or 4 arcsec diameter, respectively.

Long-slit spectra were obtained using the LRIS instrument on 01 February 1995, in good conditions.  A single 300 s exposure was obtained for the QSO component A, and two 1200 s exposures capturing the components B, C, \#3, and \#8, with a slit PA = 69$^\circ$.  We used a 300 lines mm$^{-1}$ grating, and 1.5 arcsec wide slit, giving a mean dispersion of $\sim 2.49$ \AA\ pixel$^{-1}$ and spectroscopic resolution of FWHM $\sim 17$ \AA, with a wavelength coverage from $\sim 3850$ to $\sim 8890$ \AA.  The data were reduced using standard techniques.  The spectra of the components A, B, and C are shown in Fig. 3; it is evident that C is a QSO component at the same redshift as A and B.  No redshifts could be obtained for the faint objects (probably galaxies) \#3 and \#8, but they show no indications of AGN in their spectra.

The J2000 coordinates of the QSO A, taken as a mean of the USNO A2 and B1 catalogs, are
$\alpha ~=~ 14^h ~32^m ~29.232^s$,
$\delta ~=~ -01^\circ ~06^\prime ~15.68^{\prime\prime}$.
The offsets to the component B are
$\Delta \alpha ~=~ 4.46^{\prime\prime}$ W and
$\Delta \delta ~=~ 2.49^{\prime\prime}$ N,
and to the component C are
$\Delta \alpha ~=~ 1.22^{\prime\prime}$ W and
$\Delta \delta ~=~ 4.11^{\prime\prime}$ N,
taken from \cite{fau03}.
They correspond to the projected separations A-B, A-C, and B-C of 5.11, 4.29, and 3.62 arcsec respectively, or 43, 36, and 30 proper kpc (appropriate for the gravitationally bound systems), or 131, 110, and 93 comoving kpc.

\cite{fau03} give the magnitudes for the QSO A: $R = 17.45$, $J = 16.32$, $H_{F160W} = 15.88$, and $K_s = 14.94$ mag, with errors of 0.01 mag.
From the SDSS DR4 data server, we get $u = 17.88$, $g = 17.81$, $r = 17.68$, $i = 17.48$, and $z = 17.15$ mag.
For the magnitude differences between the brightest two components, \cite{fau03} give
$\Delta R _{B-A} = 3.40 \pm 0.02$,
$\Delta J _{B-A} = 3.68 \pm 0.01$,
$\Delta H _{B-A} = 3.39 \pm 0.01$, and
$\Delta K _{B-A} = 3.62 \pm 0.01$ mag,
i.e., modest, but statistically significant differences.
\cite{hen06} give $\Delta i _{B-A} = 3.34$ mag, and
\cite{mye06} give $\Delta u _{B-A} = 3.52$ and
$\Delta g _{B-A} = 3.27$ mag.

While the optical flux ratio A/B is $25 \pm 3$, in the X-rays this flux ratio is 
significantly smaller, $5.3 \pm 1.8$, on the basis of ChaMP measurements \citep{kim06}.  We also note that the QSO A is detected by FIRST at $S_{1.4 GHz} = 0.93 \pm 0.16$ mJy, but QSO B is below their detection limit; deeper radio observations would be very useful.

Similarly, for the magnitude differences between the components A and C \cite{fau03} give
$\Delta R _{C-A} = 6.08 \pm 0.08$,
$\Delta J _{C-A} = 5.34 \pm 0.08$,
$\Delta H _{C-A} = 5.29 \pm 0.05$, and
$\Delta K _{C-A} = 5.34 \pm 0.16$ mag.
QSO A is saturated in the Keck $I$ band images, but measure $\Delta I _{C-B} = 2.85 \pm 0.2$ mag.  Assuming $\Delta I _{B-A} \approx \Delta i _{B-A} = 3.34$ mag, we infer $\Delta I _{C-A} \approx 6.2 \pm 0.2$ mag, or, using the SDSS measurement, $i_C \approx 23.6$ mag.  Put another way, the colors of the three components from \cite{fau03} are 
$(R-K)_A = 2.49 \pm 0.03$, 
$(R-K)_B = 2.27 \pm 0.03$, and 
$(R-K)_C = 3.23 \pm 0.21$; and 
$(R-J)_A = 1.13 \pm 0.03$, 
$(R-J)_B = 0.85 \pm 0.03$, and 
$(R-J)_C = 1.87 \pm 0.13$.
We infer, using SDSS $i$ band and our Keck $I$ band measurements
$(i-K)_A = 2.44$, 
$(i-K)_B = 2.16$, 
$(i-K)_C \approx 3.3 \pm 0.3$.
Thus, the component C seems to be significantly redder in terms of the optical to IR colors than either A or B.  We note that the Balmer break is redshifted to be between the optical and IR bands, where the color difference seems to occur, so one possible interpretation of the data is that there is a significant starlight component in the measured IR flux from the component C, which is the faintest AGN of the three.

Fig. 3 also shows the ratios of the three spectra.  The spectra of the components A and B are broadly similar, and we note that \cite{mwf99} found that the observed spectroscopic differences between components A and B are typical for a random pair of QSOs at these redshifts.  However, the component C has a significantly $bluer$ continuum in the observed visible regime, reverse of the trend given by its optical to IR colors.  This essentially eliminates the possibility that it is made dimmer and redder by extinction due to a foreground galaxy.  Components B and C have clearly a different shape of the C IV 1549 line from the component A, and possibly in other broad lines as well.

We used the cross-correlation technique as implemented in {\tt iraf} to measure the restframe velocity shifts between the QSO spectra.  From the peak of the correlation function, we get 
$\Delta V _{AB} = 280 \pm 160$ km s$^{-1}$,
consistent with the original measurements by \cite{hew89}, and 
$\Delta V _{BC} = 100 \pm 400$ km s$^{-1}$,
i.e., consistent both with zero, and with the relative velocities typical for the bound or interacting galaxies.

Several prominent absorption systems are seen, including lines of the C IV doublet 1548.195, 1550.770; Al II 1670.787; Fe II 2367.591, 2382.765, 2586.650, 2600.173; and Mg II doublet 2796.352, 2803.531.  In the QSO A, the two strongest absorbers are at $z_{abs,A1} = 1.5115$ and $z_{abs,A2} = 1.6617$; in the QSO B, at $z_{abs,B1} = 1.8366$ and $z_{abs,B2} = 1.513$; and a strong C IV absorber is seen in the spectrum of the QSO C at $z_{abs,C1} = 1.840$.  Presumably absorbers A1 and B2, and B1 and C1 are caused by the same intervening galaxies.  Some of the faint galaxies numbered in Figs. 1 and 2 are probably responsible for these foreground systems, and thus would not be associated with the QSOs themselves.

We performed image deconvolutions of the data, using the MCS algorithm \citep{mcs98}.  The most reliable one, in terms of the independently constrained PSF, is using the VLT $R$ band data, shown in Fig. 4.  QSO images A and B are the only ones consistent with being unresolved, and an extended component is possible for the component C.  Its total magnitude, measured in a 2 arcsec aperture is $R_C = 23.47 \pm 0.2$ mag, and the flux is divided roughly equally between the unresolved AGN component ($R_{C,unres} = 24.3 \pm 0.2$ mag) and a resolved component, presumably a host galaxy ($R_{C,host} = 24.2 \pm 0.2$ mag).  Deconvolution also reveals a significant extended component to the SSW of QSO A, suggestive of a tidally distorted host galaxy.

While the differences in broad-band spectral energy distributions of the three QSO components could be in principle explained away in the context of gravitational lensing with the usual invocations of variability and time delays,  perhaps a more natural explanation is that the three components are physically distinct AGN.

\section{The Lens Hypothesis: Modeling}

\cite{fau03} presented the most comprehensive analysis to date for this system, while being unaware of the QSO nature of the component C.  Discovery of this new QSO component, its flux ratios relative to the components A and B, the geometry of the system, and the greater depth of the Keck+VLT data, provide significant new constraints for the lensing hypothesis, which we address here.  The geometry of the three QSO components is not very close to that of known lensed quasars.  There is only one triple QSO known to be lensed, MG~2016+112 \citep{law84}.

We investigate two possible lensing scenarios that may explain the geometry of QQQ~$1432-0106$.  Both involve $four$ quasar images, three of them being A, B and C.  The difference between the two scenarios is that in one case (hereafter L1) we assume that object D (indicated in Fig. 2) is the fourth quasar image and in the second case (hereafter L2) we assume that quasar A is in fact a narrow separation and highly magnified blend of two quasar images.  There is no object in the vicinity of QQQ~$1432-0106$, that is not immediately ruled out as an extra lensed image due to its far too unrealistic position relative to A, B, C or on the basis of our Keck spectra.

In both cases, the lensing galaxy potential is modeled as a Singular Isothermal Sphere (SIS) and an external shear $\gamma$ added (SIS+$\gamma$).  This type of model fits fairly well the image configuration of most known simple gravitationally lensed QSOs.  It offers 1 degree of freedom when fitting a quadruply imaged QSO with unknown lensing galaxy position.  Quadruple systems have 8 observational constraints (the 4 positions of the lensed QSO images) while the model has 7 parameters: the total mass of the lens, the position of the source relative to the lens (2 parameters), the amplitude $\gamma$ and PA $\theta_\gamma$ of the shear and the lensing galaxy position (2 parameters).

We computed lens models using the {\tt gravlens} software \citep{kee01}.  As the position of the lensing galaxy is completly unknown, we adopt a two-fold strategy.  First, we roughly explore the parameter space by fitting models with fixed $\theta_\gamma$ and galaxy positions, scanning the 2 arcsec-region around the barycenter of quasars A, B, C.  For each galaxy position, a broad range of $\theta_\gamma$ is explored.

In a second step, we select the model with the best $\chi^2$, as obtained in the first step, we adopt its lens center and shear PA, and we run {\tt gravlens}, with all the model parameters allowed to vary.  For the lensing scenario L1, we adopt 0.02 arcsec relative astrometric errors.  For scenario L2, we split image A into two sub-components A1 and A2, separated by 0.05 arcsec.  According to the HST images analyzed by \cite{fau03}, A is a single point source.  Their PSF subtraction would have detected any blend wider than 0.05 arcsec.  For A1 and A2, we increase artificially the error bars up to 0.2 arcsec, in order to reflect the poor astrometric constraint on the putative components of the blend.  Therefore, {\tt gravlens} is relatively free to recover the image positions within loose error boxes.

We do not get acceptable fits with either model.  In the first scenario, L1, our best model predicts lensed images typically 0.5 arcsec away from the observed positions as indicated by the large value of the reduced $\chi^2=1941$.  In addition, such a model predicts image D to be the brightest, and images B and C to have similar brightness, in a clear contradiction with the observations.  Such a quadruple configuration seems therefore very unlikely.  In the second scenario, L2, our best model has $\chi^2=74$.  Images B and C are fairly well reproduced but the 2 merging images A1-A2 are predicted to lie about 1.2 arcsec away from their observed positions, and we can thus reject this possibility as well.  Thus, simple lens models do not provide acceptable fit of the observed lensed image configuration.  Better fits to the data require to add a secondary lens, resulting in a model with zero degree of freedom.  With a simple SIS+$\gamma$ model, we therefore already reach the limit of the available constraints.  
 
In addition, we estimate the total mass of the lens inside the Einstein radius for our best model L2.  It can be expressed as a function of redshift as
$M_E= (2\ 10^{12} \times z_{\rm lens})\ M_\odot$ up to $z=$ 0.8, and as
$M_E= (5\ 10^{12} \times z_{\rm lens}$ to $2.5\ 10^{12})\ M_\odot$
for $0.8 < z < 1.4$.  The shear direction for this model is $PA=+102^{\circ}$ and its amplitude is $\gamma=0.19$.  In contrast to MG~2016+112, no lensing galaxy is seen in QQQ~$1432-0106$, down to $K=23.3$ (1$\sigma$ limit in a 4 arcsec aperture).  In MG~2016+112, the lens is at z=1.01 and has $H=18.5$ mag.  The mass of the putative lens, if placed at an overly optimistic $z=0.5$ is $M_E = 10^{12}\ M_\odot$, in a 4 arcsec aperture, using model L2.  This corresponds to a velocity dispersion of $\sigma=330$ km/s.  If the lens is at $z=1.4$ then $\sigma=580$ km/s.  If the lens is at $z=0.5$ then its absolute magnitude would be $M_{\rm B}=-23.2$ from the Faber-Jackson relation ($\sigma$ from lens model, without conversion from $\sigma_{\rm lens}$ to $\sigma_{\rm dyn}$).  If it is at $z=1.4$ then $M_{\rm B}=-25.5$.  Even without any correction for evolution, this translates to apparent magnitudes of $K=18.5$, and $K=17.1$ mag, respectively.  Clearly, presence of such lensing galaxies is strongly excluded by our deep $K$ band images, even if they were to be aligned with the brightest QSO component (which would be a very difficult geometry to arrange).

To summarize, we are unable to produce any plausible lensing model which would account for the observed geometry and intensity ratios of the QSO images, and moreover we have strong upper limits on the existence of necessary lensing galaxies or groups.  Moreover, we note that \cite{fau03} have placed strong limits to weak lensing distortions in this field, which would be expected if a massive lens was present, regardless of its obscuration.  Therefore, we conclude that this system is highly unlikely to be a case of gravitational lensing.

\section{Discussion and Conclusions}

The observed differences in the spectra and colors of the QSO components, and the great difficulty in modeling the system as a gravitational lens, strongly suggest that we are dealing with a case of a physical triple QSO.  The projected separations are typical of those seen in galaxy interactions.

Measurements of a QSO 3-point correlation function at such physical separations are currently not available, at any redshift.  Thus, we cannot compute a credible probability for such a configuration in a pure gravitational clustering regime.  However, we know that binary QSOs at high redshifts are highly overabundant relative to the predictions derived from simple galaxy and mass clustering models \citep{djo91,kfm99,hen06}.  This excess is naturally understood as a consequence of enhanced propensity for fueling of AGN in dissipative galaxy interactions.  It is then perhaps not surprising that a triple QSO system can be found in similar circumstances.

Indeed, we observe QQQ~$1432-0106$ at a redshift where the merging activity (and the comoving density of QSOs) was roughly at its peak.  Several components of the system can then be interpreted as a compact galaxy group caught in process of interacting, with AGN activity ignited simultaneously in 3 of the participating host galaxies.  Hierarchical merging in such systems could also lead to formation of triple SMBH systems, and a number of interesting effects; see, e.g., \cite{hl06} for a discussion and references.

As the census of such triple-QSO systems (and binary QSOs themselves) grows with the future observations, and the selection effects are better understood, we will have a new observational probe of the processes of galaxy-SMBH co-evolution at high redshifts.

\acknowledgments

We are grateful to the staff of the W.  M.  Keck and ESO-VLT observatories for their expert assistance in the course of our observing runs, and to M.  Pahre with assistance with some of the Keck observations.  The Keck data used in this paper were obtained in a collaborative project involving Profs.  J.  Cohen, B.T.  Soifer, R.  Blandford, and G.  Neugebauer, and I.  Smail.  SGD, AAM, and EG acknowledge a partial support from the NSF grant AST-0407448, and the Ajax Foundation.  FC, GM and DS are financially supported by the Swiss National Science Foundation (SNSF).  SGD acknowledges the hospitality of EPFL and Geneva Observatory, where some of this work was performed.


\clearpage


\begin{figure}
\plotone{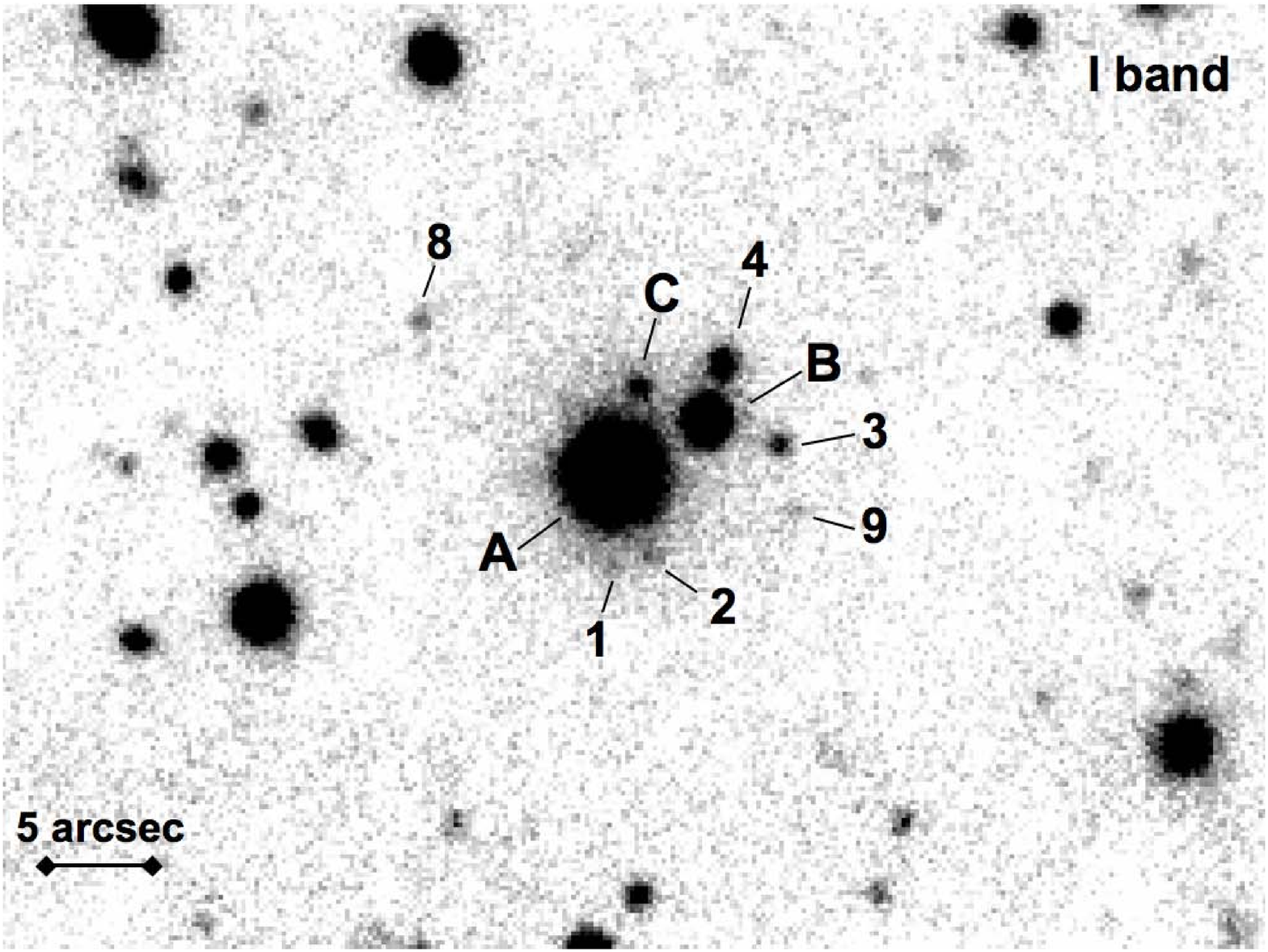}
\caption{
Keck $I$ band image of the field, with the QSO components A, B, C, and other faint objects nearby labeled; the numbers follow and extend the notation from \cite{fau03}.  N is up, E to the left, and the scale is given by the inset bar.
\label{fig1}}
\end{figure}

\clearpage


\begin{figure}
\plotone{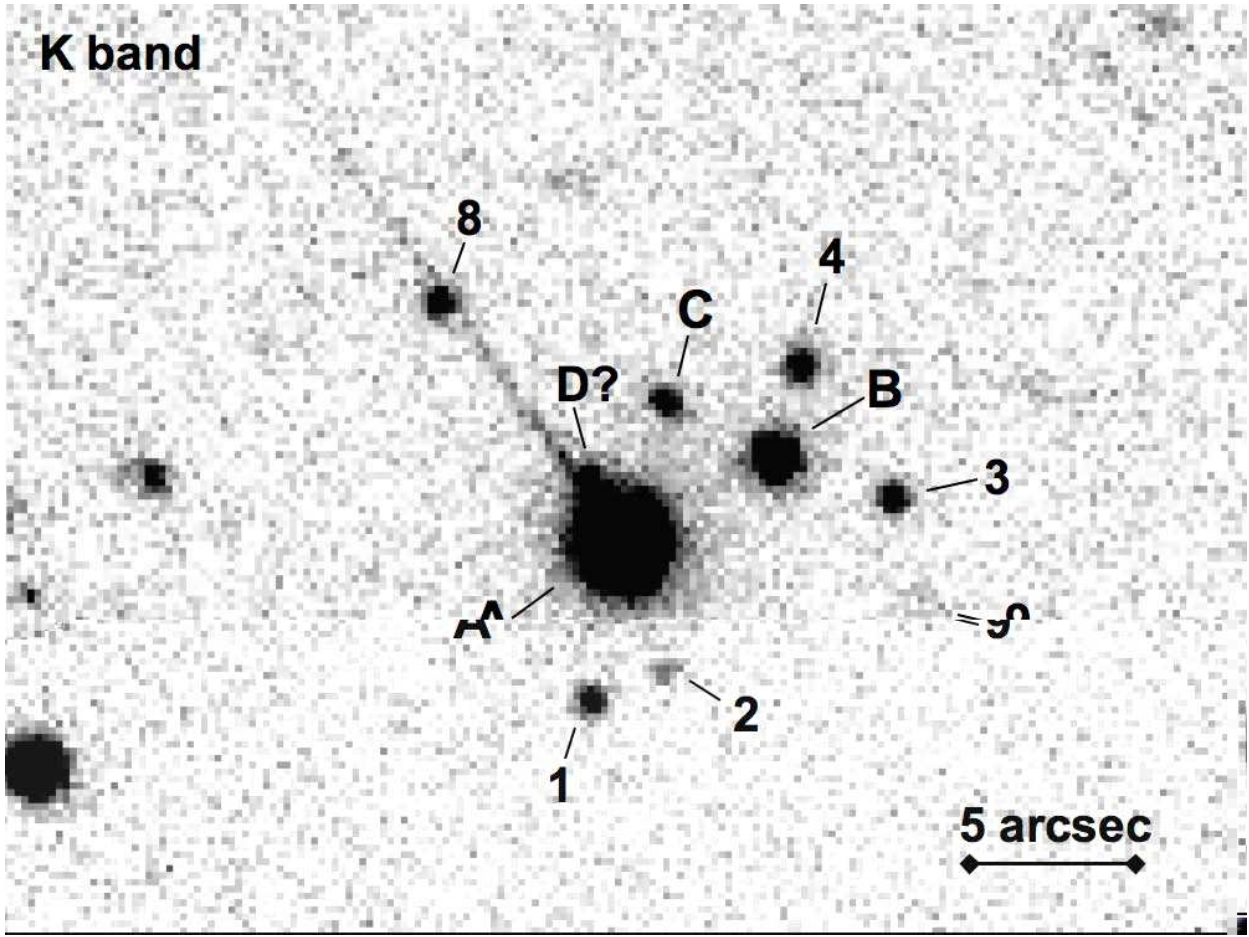}
\caption{
A sum of the Keck and VLT $K$ band images of the field.  N is up, E to the left, and the scale is given by the inset bar. The notation is the same as in Fig. 1, with a possible additional object (D?) adjacent to the QSO image A.  It is unfortunately covered by the charge transfer artifact (diagonal streak) from the Keck NIRC images, prior to the rotation to a standard cardinal orientation.
\label{fig2}}
\end{figure}

\clearpage


\begin{figure}
\epsscale{.7}
\plotone{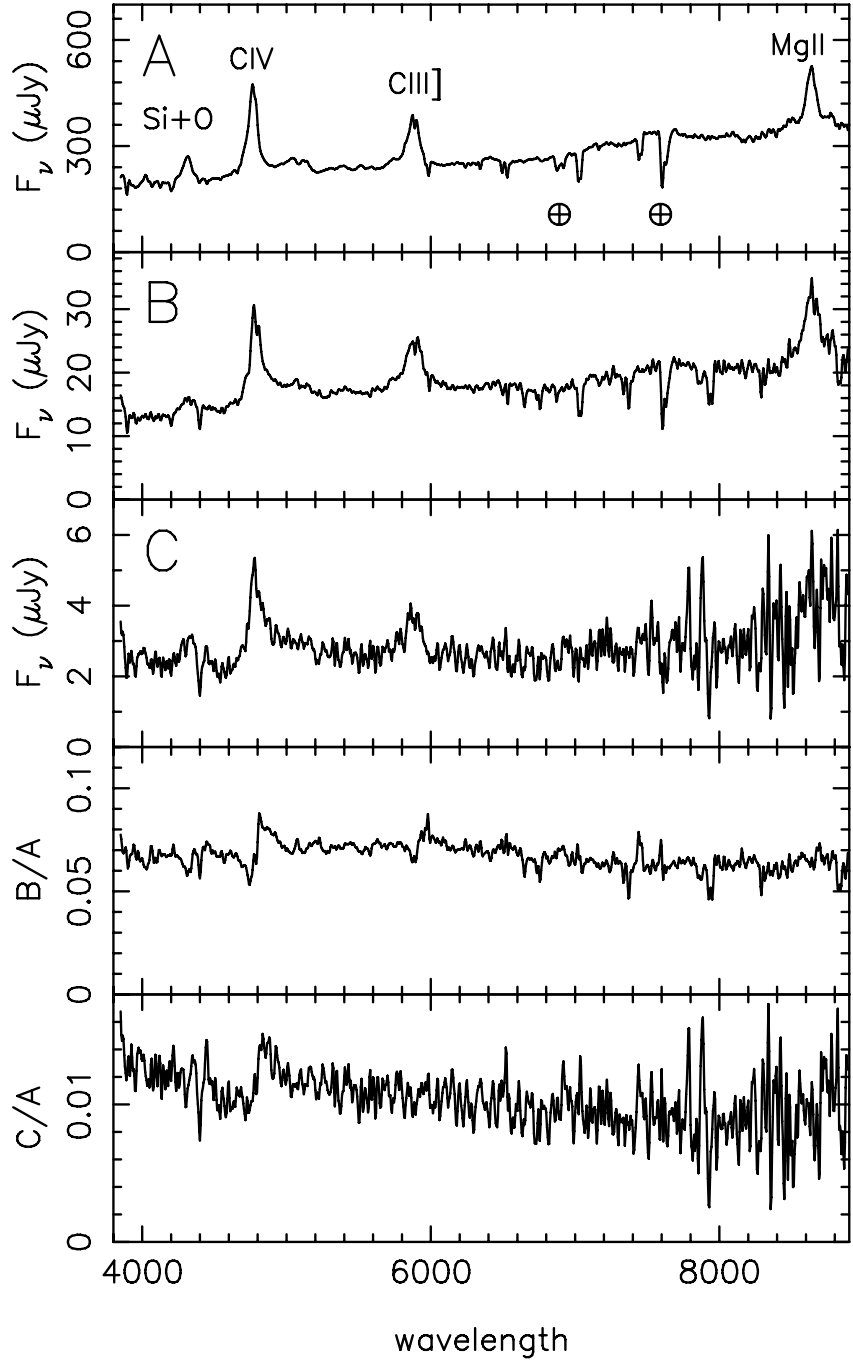}
\caption{
Spectra of the QSO components A, B, and C (the top 3 panels), and their ratios (the bottom 2 panels), obtained at the Keck.  The prominent emission lines, Si IV + O IV 1400, C IV 1549, C III] 1909, and Mg II 2799, are labeled, as well as the atmospheric absorption A and B bands.  A number of metallic absorption systems are detected; see the text for more details.
\label{fig3}}
\end{figure}

\clearpage


\begin{figure}
\epsscale{1.00}
\plotone{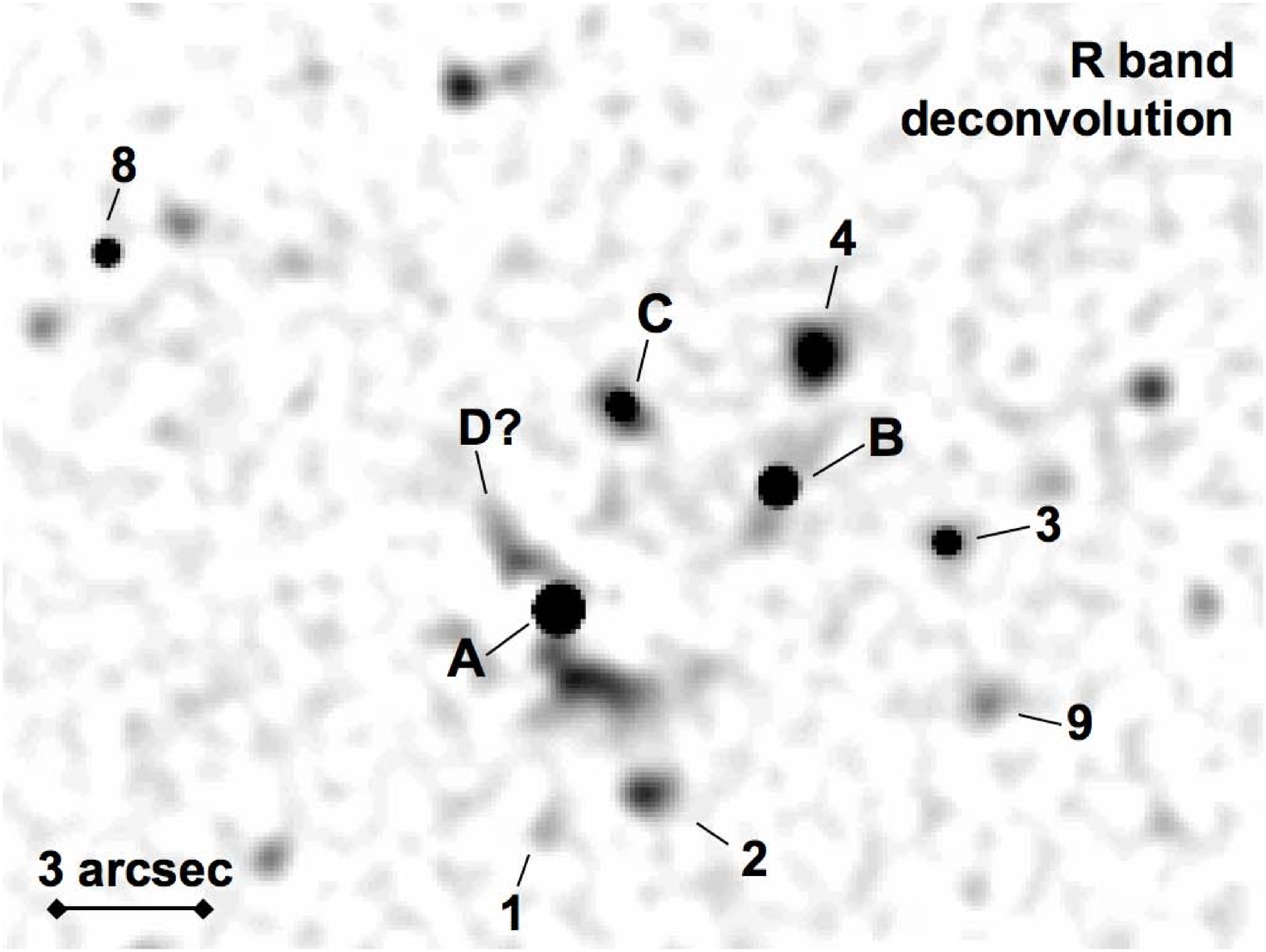}
\caption{
Image deconvolution of the VLT $R$ band image. The notation is the same as in Fig. 2.  Note the extended, resolved structure SSW of the QSO A, suggestive of a tidally distorted galaxy; the possible object D may be a part of it.  Hints of the resolved host galaxies are also seen underlying QSOs B and C.
\label{fig4}}
\end{figure}

\clearpage


\end{document}